\documentclass[aps,pra,reprint,showpacs,groupedaddress]{revtex4-1}
\usepackage[dvips]{graphicx}
\usepackage[dvips,colorlinks=true,bookmarks=false,citecolor=blue,urlcolor=blue]{hyperref} %latex w/dvips
\usepackage{mathtools}

\begin{document}
\title{Quantum multi-parameter estimation with generalized balanced multi-mode NOON-like states}

\author{Lu \surname{Zhang}}
\email[Email: ]{lu@ou.edu}
\author{Kam Wai Clifford \surname{Chan}}

\affiliation{School of Electrical and Computer Engineering, University of Oklahoma--Tulsa, Tulsa, Oklahoma 74135, USA}

\begin{abstract}
The simultaneous multi-parameter estimation problem using a class of multi-mode entangled states is investigated in this paper.  Specifically, the problem of optical phase imaging is considered and the quantum probe is taken to be a balanced coherent superposition of components with an arbitrary quantum state in one mode and vacuum states in all the other modes, which is a generalization of the multi-mode NOON state.  The analytical form for the quantum Cram\'er-Rao bound (QCRB) is presented, which shows the performance by providing a lower bound of the estimation uncertainty.  It is shown that the NOON state has the worst performance among those in the class of the entangled states considered.  We also analyze in detail four different scenarios, which are the NOON state, the entangled coherent state, the entangled squeezed coherent state, and the entangled squeezed vacuum state.  From the comparison among these four states, we find that when the mean photon number is fixed, the squeezed vacuum state has the smallest QCRB, followed by the squeezed coherent state, entangled coherent state, and NOON state.  We also illustrate that the balanced entangled state can perform better than a more generalized unbalanced form studied in previous works for certain scenarios.  Finally, we give an experimental setup for producing a two-mode entangled state that can beat the NOON state in quantum metrology.
\end{abstract}
\date{\today}

\maketitle

%%%%%%%%%%%%%%%%%%%%%%%%%%%%%%%%%%%%%%%%%%%%%%%%%%%%%%%%%%%%%%%%%%%%%%%%%%%%%%%%%%%%%%%%%%%%
\section{Introduction}
Quantum metrology, also known as quantum parameter estimation, aims at studying the ultimate measurement precision of physical parameters that is limited by the laws of quantum theory~\cite{Helstrom1976_Book_quantumdetectionandestimationtheory,Giovannetti2004_entanglement/squeezing_Heisenberg_single}.  In general, the quantum precision depends on the properties of the probing quantum states, the interaction of the states with the target, and the strategy of measurement.  For quantification, the quantum Cram\'er-Rao bound (QCRB)~\cite{Helstrom1976_Book_quantumdetectionandestimationtheory,Braunstein1994_statisticalDistance}, which sets the lower bound of the estimation uncertainty with any possible measurement strategy, is customarily employed. Under this circumstance, the goal of the studies on quantum parameter estimation is to reduce the QCRB and hence achieve super-sensitivity. Using classical light sources, the QCRB can reach the standard quantum limit (SQL) of precision with a scaling $1/\sqrt{N}$, in which $N$ is the mean photon number of the quantum probe.  On the other hand, by taking advantages of quantum properties such as entanglement and squeezing~\cite{Giovannetti2004_entanglement/squeezing_Heisenberg_single}, one can beat the SQL and approach the Heisenberg limit with a scaling $1/N$, which has a $\sqrt{N}$ benefit over the classical counterpart.

Extensive theoretical and experimental research work has been conducted for the estimation problem of a single parameter~\cite{Bollinger1996_single_NOON, gerry2001_single_ECS, Kok2004_single_NOON_l,Giovannetti2006_entanglement_eigenstates_single, Dorner2009_single_NOON_noise, Rivas2010_single_CS_lnl, Joo2011_singe_ECS_l, krischek2011_experiments_single, Datta2011_single_HBstate, Joo2012_singe_ECS_nl, Tan2014_single_MZI_squeezeThermal_evenOdd, Lee2015_single_multiheadcat_l, Liu2016_multi_GeneralizedECS_lnl, Knott2016_single_squeezed}. In particular, the NOON state~\cite{Boto-etal2000, Lee2002_single_NOON_name} is customarily considered to show the ability of breaking the SQL~\cite{Bollinger1996_single_NOON, Kok2004_single_NOON_l, Dorner2009_single_NOON_noise}.  On the other hand, there has been recently increasing interest in the study of multi-parameter estimation in view of the potential advantage of increased estimation efficiency in simultaneous multiple parameter determination.  In cases where there are couplings among the different parameters or the underlying generators are non-commutative, it was found that quantum entanglement may not necessarily be advantageous~\cite{Kok-etal2015,Baumgratz-Datta2016}.
For the other class of problems where the parameters do not couple to each other, certain entangled states have been studied in order to reach the Heisenberg limit.
Especially, the latter situation is applicable to optical phase imaging in which the unknown parameters represent the phase shifts at different spatial pixels induced by a phase object.  In such context, Humphreys \textit{et al.}~\cite{Humphreys2013_multi_NOON_l} showed that the multi-parameter estimation using a multi-mode NOON state can approach the Heisenberg limit with an $O(d)$ advantage over the independent estimation of $d$ phase shifts using $d$ copies of two-mode NOON states. It should be noted that such an advantage is due to the large photon number variance in the single modes of the multi-mode state and can be reproduced using only local strategies~\cite{Knott2016_modeseparablestate}.
Aside from the NOON state, other quantum probes have also been considered, such as the cat state \cite{Lee2015_single_multiheadcat_l}, the Holland-Bernett state~\cite{Datta2011_single_HBstate}, entangled coherent state~\cite{Ono2010_multi_ECS_loss, Joo2011_singe_ECS_l, Joo2012_singe_ECS_nl, Cheng2014_2_NOON_ECS_lnl_simultaneous, Liu2013_multi_ECS_loss, Liu2016_multi_GeneralizedECS_lnl, Zhang2013_multi_ECS_loss, Jing2014_multi_ECS_loss}, Gaussian state~\cite{Pinel2013_multi_Gaussian, Safranek2015_multi_gaussian}, etc.

Besides the quantum probe, another significant factor determining the ultimate precision in quantum imaging is how the state interacts with the target object, which is usually represented by a unitary operator~\cite{Liu2015_unitary_QFI_H_mixed_lnl}. For the estimation of a single unknown parameter $\theta$, the interaction operator can be written as $U_\theta=e^{i\theta \cdot \hat{H}}$~\cite{Joo2011_singe_ECS_l}, with $\hat{H}=a^\dagger a$ linear in the photon number operator. This linear characterization is more common in experiments~\cite{kacprowicz2010_experiments_single, giovannetti2011_experiments_single, krischek2011_experiments_single} and hence more often considered than the nonlinear cases with $\hat{H}=(a^\dagger a)^k$ ($k\geq 2$), even though the latter can potentially reduce the uncertainty with a degree $k$ and reach $1/N^{k}$~\cite{Joo2012_singe_ECS_nl}.

In this paper, we presented a class of quantum entangled states for the study of multi-parameter quantum metrology in optical phase imaging. It is generalized from the multi-mode NOON state, where the non-vacuum mode can have arbitrary photon statistics instead of a fixed number of photons (Fock state). We obtained an analytical form of the QCRB using this balanced state, i.e., the probability amplitude of the reference component is the same as that of the components probing the quantum system. Specifically, we studied the imaging performances when the probe state is a multi-mode NOON state, entangled coherent state, entangled squeezed coherent state, and entangled squeezed vacuum state.  Comparisons were made among these different probe states and it was proven that, with the same mean number of photons, the entangled squeezed vacuum state has the lowest estimation uncertainty and hence leads to the best imaging precision, followed by squeezed coherent state, entangled coherent state, and NOON state. It is also interesting to notice that for the squeezed coherent state scenario, the higher the squeezing degree is, the lower the uncertainty will be.  In addition, we showed that the balanced entangled state can perform better than its unbalanced counterpart adopted in the previous works~\cite{Liu2016_multi_GeneralizedECS_lnl, Humphreys2013_multi_NOON_l} under the squeezed vacuum scenario. Finally, we described an experimental setup that can produce a two-mode entangled state, which has the ability to beat NOON state with respect to the ultimate sensitivity.

%%%%%%%%%%%%%%%%%%%%%%%%%%%%%%%%%%%%%%%%%%%%%%%%%%%%%%%%%%%%%%%%%%%%%%%%%%%%%%%%%%%%%%%%%%%%
\section{Quantum Cram\'er-Rao bound for multi-parameter estimation using balanced entangled state}
\begin{figure}[!t]
    \centering\includegraphics[width=7cm]{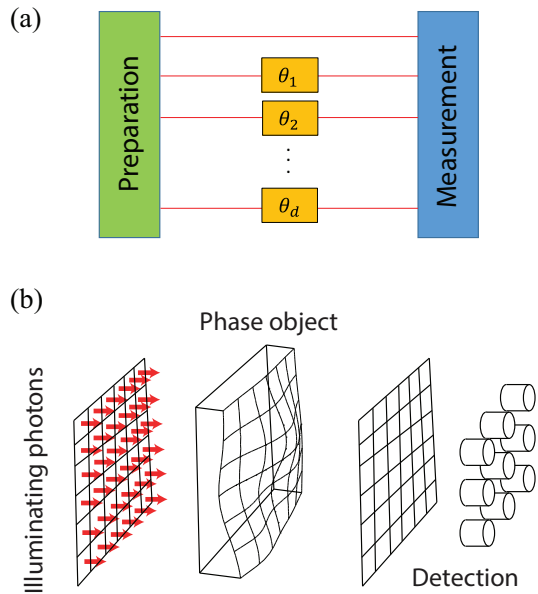}
    \caption{(a) Schematic of multiple-phase estimation with $d$ parameters. (b) Discretized phase imaging model.}
    \label{fig:schematic}
\end{figure}
In the following, we study the simultaneous estimation problem for $d$ independent phases. The schematic of the setup follows~\cite{Humphreys2013_multi_NOON_l} and is depicted in Fig.~\ref{fig:schematic}. The quantum probe we choose is a balanced $(d+1)$-mode entangled pure state, which can be written as
\begin{equation}
    |\Psi\rangle
    =
    b\sum_{m=0}^{d}|0\rangle_0 |0\rangle_1 |0\rangle_2 \cdots |\psi\rangle_m \cdots |0\rangle_d.
    \label{eq_generalized_state}
\end{equation}
It is a superposition of $d+1$ multi-mode quantum states, each of which has arbitrary photon statistics in mode $m$ denoted by state $|\psi\rangle_m$ and no photons in the other modes. The normalization coefficient is given by
\begin{equation}
    b=\frac{1}{\sqrt{d+1}\sqrt{1+d|\langle \psi|0\rangle|^2}}.
\end{equation}
By convention, the $m=0$ mode is chosen as the reference.  Note that the state has balanced weights between the reference mode and the other information modes, which will be shown below to perform better than the unbalanced case under certain circumstance.
The total mean photon number for this state is
\begin{equation}
    \bar{n} \equiv \langle\Psi| \left(\sum_{m=0}^d a_m^\dagger a_m\right) |\Psi\rangle
    = \frac{\tilde{n}}{1+d\left|\langle\psi|0\rangle\right|^2},
\label{eq_mean_photon_number}
\end{equation}
where $\tilde{n}=\langle\psi|a^\dagger a|\psi\rangle$ is the mean photon number for the single mode state $|\psi\rangle$. It is noticeable that $\bar{n} \leq \tilde{n}$ is always true, which is due to a finite probability of state $|\psi\rangle$ containing no photons.  The equality is satisfied only when $|\langle\psi|0\rangle|^2=0$ as in the case, for example, of a NOON state.
In practical sensing problems, $\bar{n}$ may be more meaningful than $\tilde{n}$ since it characterizes on average how many photons are used for the simultaneous parameter estimation.

In this paper, we assume that the phase object induces linear phase shifts to the probe. Therefore the unitary operator is denoted as
\begin{eqnarray}
    U_{\boldsymbol{\theta}}
    &=&
    \text{exp}\left( i \boldsymbol{\theta} \cdot \boldsymbol{\hat{H}} \right) \\
    &=&
    \text{exp}\left(i \sum_{m=1}^{d} \theta_m \hat{H}_{m}\right)=\prod_{m=1}^{d}\text{exp}\left(i \theta_m \hat{H}_{m}\right),
\label{eq_unitary_operator}
\end{eqnarray}
where $\boldsymbol{\theta}=(\theta_1,\theta_2...,\theta_d)$ represents $d$ independent phases and $\hat{H}_{m}=a_m^\dagger a_m$ is the photon number operator for the mode $m$.
In writing Eq.~(\ref{eq_unitary_operator}), we have assumed that the reference mode has a phase set to zero, which practically can be achieved by phase-locking the reference arm with some external reference~\cite{Jarzyna-etal2012}. The inclusion of the reference mode is to make the comparison with the multi-mode states considered in the previous works more explicit~\cite{Liu2016_multi_GeneralizedECS_lnl, Humphreys2013_multi_NOON_l}.
The output state after the propagation process then reads as $|\Psi_{\boldsymbol{\theta}}\rangle=U_{\boldsymbol{\theta}}|\Psi\rangle$.

Given the probe state and the evolution operator, we can now calculate the quantum Cram\'er-Rao bound, which is inversely proportional to the quantum Fisher information.  Then, we get the lower bound of the estimation uncertainty
\begin{equation}
    |\delta\boldsymbol{\theta}|^2
    \geq
    |\delta\boldsymbol{\theta}|_{\text{QCRB}}^2
    =
    \text{Tr}(I_{\boldsymbol{\theta}}^{-1}),
\end{equation}
where $I_{\boldsymbol{\theta}}$ is the $d\times d$ quantum Fisher information matrix.
The condition for the QCRB being saturated is $\text{Im}\langle\Psi_{\boldsymbol{\theta}}| L_l L_m |\Psi_{\boldsymbol{\theta}}\rangle=0$~\cite{Matsumoto2002_QCRB_condition_pure}, where $L_l=2\left(|\partial_l\Psi_{\boldsymbol{\theta}}\rangle\langle\Psi_{\boldsymbol{\theta}}| + |\Psi_{\boldsymbol{\theta}}\rangle \langle\partial_l \Psi_{\boldsymbol{\theta}}|\right)$ is the symmetric logarithmic derivative (SLD) and $\partial_l$ denotes the partial derivative with respect to $\theta_l$. Since $\text{Re}\left(\langle\Psi_{\boldsymbol{\theta}}|\partial_l\Psi_{\boldsymbol{\theta}}\rangle\right)=0$ and $\text{Im}\left(\langle\partial_m\Psi_{\boldsymbol{\theta}}|\partial_l\Psi_{\boldsymbol{\theta}}\rangle\right)=0$, it is straightforward to find out that the saturation condition is always true for our scenario.
Finally, by calculating the quantum Fisher information matrix as~\cite{Paris2009_QFIMatrix}
\begin{equation}
    I_{\boldsymbol{\theta}}=4 b^2\langle \hat{H}^2 \rangle I - 4 b^4 \langle \hat{H} \rangle^2 O,
\end{equation}
where $\langle...\rangle$ denotes $\langle\psi|...|\psi\rangle$ with the mode number index $m$ in $\hat{H}$ omitted for simplicity, and $I$ and $O$ respectively represent the $d\times d$ identity matrix and the matrix with all elements equal to 1, we obtain the expression of the quantum Cram\'{e}r-Rao lower bound,
\begin{equation}
    |\delta\boldsymbol{\theta}|_{\text{QCRB}}^2
    =
    \frac{d}{4 \langle \hat{H}^2 \rangle}
    \left(\frac{1}{b^2}+\frac{1}{R - b^2 d}\right) ,
    \label{eq_QCRB_function}
\end{equation}
where $R \equiv \langle \hat{H}^2 \rangle / \langle \hat{H} \rangle^2$.

Equation~(\ref{eq_QCRB_function}) gives the analytical form of the QCRB for any quantum probe with the form as in Eq.~(\ref{eq_generalized_state}).  It should be noted that $|\delta\boldsymbol{\theta}|_{\text{QCRB}}^2$ is strictly positive here, for if it were zero, the required condition would be
\begin{equation}
    R = (d-1)b^2
    = \left(\frac{d-1}{d+1}\right) \frac{1}{1+d|\langle \psi|0\rangle|^2} < 1 ,
\end{equation}
which contradicts the nonnegativity of the variance $\langle \hat{H}^2 \rangle - \langle \hat{H} \rangle^2$.

It is remarked, as also mentioned previously, that local strategies can reproduce the performance of the multi-mode quantum states~\cite{Knott2016_modeseparablestate}.  The corresponding mode-separable state for Eq.~(\ref{eq_generalized_state}) is
$
    |\Psi\rangle_\text{LS}
    =
    \mathcal{N} \left(|\psi\rangle + \nu |0\rangle\right)^{\otimes d}
$,
where $\mathcal{N}$ is the normalization.  This local mode-separable state can perform similarly to the multi-mode state~(\ref{eq_generalized_state}) in phase estimation when $\nu \propto \sqrt{d}$~\cite{Knott2016_modeseparablestate}.  It has the advantages of being easier to create and more resilient to loss, despite of its large vacuum component that makes it increasing inefficient with a larger $d$.  Since the main focus of this paper is to study the properties of the multi-mode NOON-like states, the mode-separable forms will not be elaborated in the subsequent discussion.

%%%%%%%%%%%%%%%%%%%%%%%%%%%%%%%%%%%%%%%%%%%%%%%%%%%%%%%%%%%%%%%%%%%%%%%%%%%%%%%%%%
\section{Analysis of the Lower Bound of Uncertainty for Selected Scenarios}
In this section, we consider four specific scenarios, which are the multi-mode NOON state $|\Psi\rangle_N$ ($|\psi\rangle=|N\rangle$)~\cite{Humphreys2013_multi_NOON_l}, the entangled coherent state (ECS) $|\Psi\rangle_\text{c}$ ($|\psi\rangle=|\alpha\rangle$)~\cite{Liu2016_multi_GeneralizedECS_lnl}, the entangled squeezed vacuum state (ESVS) $|\Psi\rangle_\text{sv}$ ($|\psi\rangle=|r\rangle$), and the entangled squeezed coherent state (ESCS) $|\Psi\rangle_\text{sc}$ ($|\psi\rangle=|\alpha',r'\rangle$)~\cite{Agarwal2013}.
Note that the two-mode ESVS has been studied in~\cite{Knott2016_single_squeezed, Lee2016_2mode_ESVS}.
Without loss of generality, we assume the amplitudes $\alpha$ and $\alpha'$, and the squeeze factors $r$ and $r'$ are real numbers.
The mean photon numbers for the balanced multi-mode entangled states above are
\begin{eqnarray}
    &\displaystyle
    \bar{n}_N = \tilde{n}_N ,
\quad
    \bar{n}_\text{c} = \frac{\tilde{n}_\text{c}}{1+d e^{-\alpha^2}} ,
\quad
    \bar{n}_\text{sv} = \frac{\tilde{n}_\text{sv}}{1+d /\cosh{r}} ,
    &
\nonumber\\
    &\displaystyle
    \bar{n}_\text{sc} = \frac{\tilde{n}_\text{sc}}{1+d e^{-\alpha'^2\left(1-\tanh r'\right)} /\cosh{r'}},
    &
    \label{eq_mean_photon_number_4}
\end{eqnarray}
where $\tilde{n}_N = N$, $\tilde{n}_\text{c} = \alpha^2$, $\tilde{n}_\text{sv} = \sinh^2{r}$, and $\tilde{n}_\text{sc} = \alpha'^2+\sinh^2{r'}$.

For the balanced multi-mode NOON state, its minimum uncertainty takes a simple form:
\begin{equation}
    |\delta\boldsymbol{\theta}|_{N}^2
    =
    \frac{d(d+1)}{2\bar{n}_N^{2}}
    =
    \frac{d(d+1)}{2\tilde{n}_N^{2}}
    =
    \frac{d(d+1)}{2N^{2}},
    \label{eq_QCRB_NOON}
\end{equation}
where the number of phases $d$ and the photon number $N$ are independent parameters.

To compare the QCRB with respect to different quantum probes, we first need to set the different probes under the same footing.  A natural choice is the mean total photon number $\bar{n}$ given by Eq.~(\ref{eq_mean_photon_number}). Under this circumstance, we rewrite Eq.~(\ref{eq_QCRB_function}) as a function of a defined factor $f \equiv \langle\hat{H}\rangle / \langle \hat{H}^2 \rangle$:
\begin{equation}
    |\delta\boldsymbol{\theta}|_{\text{QCRB}}^2
    =
    \frac{d(d+1)}{4} f
    \left(\frac{1}{\bar{n}} + \frac{1}{ \frac{(d+1)}{f} - d \bar{n} }\right).
    \label{eq_QCRB_function_f}
\end{equation}
It is remarked that $f$ is related solely to the expectations with respect to the single mode state $|\psi\rangle$, whereas $d$ and $\bar{n}$ are the features of the entire multi-mode state $|\Psi\rangle$.
When $d$ and $\bar{n}$ are fixed, $|\delta\boldsymbol{\theta}|_{\text{QCRB}}^2$ is a monotonic increasing function of $f$, where $0 < f \leq 1/\bar{n}$ and the equality for the upper bound holds for $\langle \hat{H}^2 \rangle = \langle \hat{H}\rangle^2$ and $|\langle \psi|0\rangle|^2 = 0$, i.e., NOON state.
In fact, it can be shown from Eq.~(\ref{eq_QCRB_function_f}) that
\begin{equation}
    |\delta\boldsymbol{\theta}|_{\text{QCRB}}^2 \le \frac{d(d+1)}{2\bar{n}^2}
    \label{eq_QCRB_ineq}.
\end{equation}
Therefore, any entangled state $|\Psi\rangle$ with $\langle \hat{H}^2 \rangle > \langle \hat{H}\rangle^2$
can achieve a lower estimation uncertainty than the NOON state (i.e., $|\delta\boldsymbol{\theta}|_{\text{QCRB}}^2 \leq |\delta\boldsymbol{\theta}|_{N}^2$). Specifically, when $d$ and $\bar{n}$ are fixed (i.e., $\bar{n}_N = \bar{n}_\text{c} = \bar{n}_\text{sc} = \bar{n}_\text{sv}$) for the four scenarios mentioned at the beginning of this section, we can get that the mean photon numbers of their non-vacuum mode $|\psi\rangle$ satisfy $\tilde{n}_N < \tilde{n}_{\text{c}} < \tilde{n}_{\text{sc}} < \tilde{n}_{\text{sv}}$, and hence we can derive $f_N > f_{\text{c}} > f_{\text{sc}} > f_{\text{sv}}$ (see Appendix A for details), which leads to the relations of the QCRB for the four specific cases as
\begin{equation}
    |\delta\boldsymbol{\theta}|_{N}^2 > |\delta\boldsymbol{\theta}|_{\text{c}}^2 > |\delta\boldsymbol{\theta}|_{\text{sc}}^2 > |\delta\boldsymbol{\theta}|_{\text{sv}}^2.
    \label{eq_QCRB_comparison_noon_coh_sc_sv}
\end{equation}
As we can see, the entangled squeezed vacuum state has the lowest uncertainty, followed by entangled squeezed coherent state, entangled coherent state, and NOON state. This suggests that with the same number of photons, entangled squeezed vacuum state can reach the highest sensitivity in quantum metrology.

\begin{figure}[!t]
\centering
\includegraphics[width=8cm]{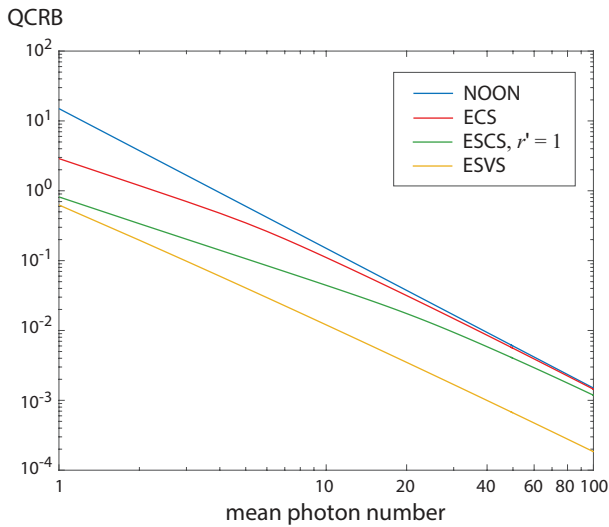}
\caption{Plots of the QCRB for the NOON state (blue), ECS (red), ESCS with $r'=1$ (green), and ESVS (yellow) with respect to the mean photon number $\bar{n}$ using Eqs.~(\ref{eq_QCRB_function_f}) and~(\ref{4f}).
The number of phases is taken to be $5$. For the NOON state, the discrete points are interpolated to give better visualization.
}
\label{Fig1_QCRB_4cases}
\end{figure}

\begin{figure}[!t]
\centering
\includegraphics[width=8cm]{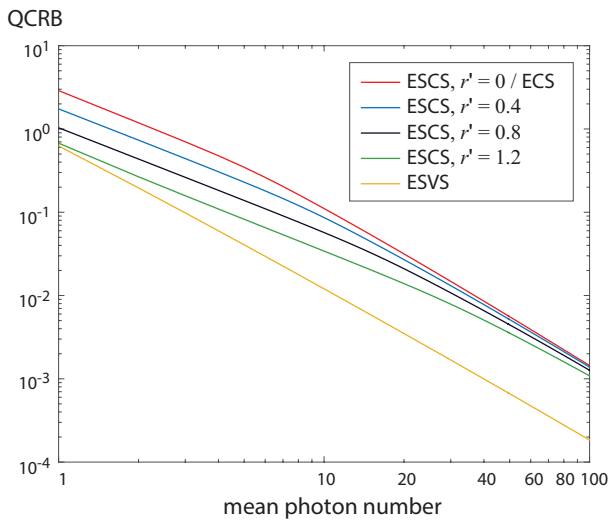}
\caption{Plots of the QCRB for the entangled squeezed coherent states (ESCS) with different squeeze factors $r'=0.4$ (blue), $r'=0.8$ (black), and $r'=1.2$ (green). The red curve and the yellow curve correspond to the ECS and ESVS cases as in Fig.~\ref{Fig1_QCRB_4cases}, which set the upper and lower boundaries for ESCS QCRB.
}
\label{Fig2_QCRB_sc_differentR}
\end{figure}

To illustrate the inequality in Eq.~(\ref{eq_QCRB_comparison_noon_coh_sc_sv}) explicitly and more exactly, we plot the QCRB for the four scenarios with respect to the mean total photon number $\bar{n}$ under the condition of $d=5$ phases, as shown in Fig.~\ref{Fig1_QCRB_4cases}. Since the mean photon number of the squeezed coherent state depends on both $\alpha'$ and $r'$, we fix its squeeze factor $r'=1$ for proper comparison.  Note that Fig.~\ref{Fig1_QCRB_4cases} confirms the observation in the previous paragraph.  Moreover, the ESVS QCRB is an order of magnitude smaller than the NOON QCRB for a wide range of the mean photon number.

It is remarked that the entangled squeezed coherent states with the same mean photon number but different combinations of $\alpha'$ and $r'$ display different performances.  Generally speaking, it can achieve lower uncertainty with a larger squeeze factor $r'$, which has been shown in Fig.~\ref{Fig2_QCRB_sc_differentR}.
From the numerical calculations, we see that as we increase $r'$ from the value $r'=0$ (red), the curve approaches the curve for squeezed vacuum state (yellow). The higher the squeeze factor is, the lower its QCRB can be saturated.

%%%%%%%%%%%%%%%%%%%%%%%%%%%%%%%%%%%%%%%%%%%%%%%%%%%%%%%%%%%%%%%%%%%%%%%%%%%%%%%%%%%%%%%%%%%%%%%%%%%%%%%%%
\section{Comparison between the Balanced and Unbalanced Cases}

In this section, we compare the performances of our presented balanced entangled state against an unbalanced state as used in~\cite{Humphreys2013_multi_NOON_l,Liu2016_multi_GeneralizedECS_lnl} with the form
\begin{equation}
    |\Psi\rangle_\text{unb}
    =
    c |\psi\rangle_0 |0\rangle_1  \cdots |0\rangle_d
    +
    b\sum_{m=1}^{d}|0\rangle_0 \cdots |\psi\rangle_m \cdots |0\rangle_d  ,
    \label{eq_unbalanced_state}
\end{equation}
where the weights $b$ and $c$ satisfy
\begin{equation}
    A b^2 + B bc +c^2 =1,
    \label{eq_unbalanced_bccondition}
\end{equation}
with $A=d+d(d-1)|\langle\psi|0\rangle|^2$ and $B=2d|\langle\psi|0\rangle|^2$.

The main idea of those papers is to minimize the uncertainty $|\delta\boldsymbol{\theta}|_{\text{QCRB}}^2$, which has the same form as in Eq.~(\ref{eq_QCRB_function}) but a different $b$ value, by choosing the optimal coefficients $b$ and $c$.  According to the geometric property of the ellipse of Eq.~(\ref{eq_unbalanced_bccondition}) with respect to $b$ and $c$, we can obtain the numerical range of $b^2$
\begin{equation}
    b^2\leq b_\text{bo}^2 \equiv \frac{1}{d\left(1+d|\langle\psi|0\rangle|^2\right)\left(1-|\langle\psi|0\rangle|^2\right)}.
\end{equation}
By treating $|\delta\boldsymbol{\theta}|_{\text{QCRB}}^2$ in Eq.~(\ref{eq_QCRB_function}) as a function of $b^2$, it is found that its minimum value is either achieved at the stationary point $b_\text{opt}^2 = R/(d+\sqrt{d})$ when $b_\text{opt}^2 \leq b_\text{bo}^2 $, or at the boundary point $b_\text{bo}^2$ otherwise.

\begin{figure}[!t]
\centering
\includegraphics[width=8cm]{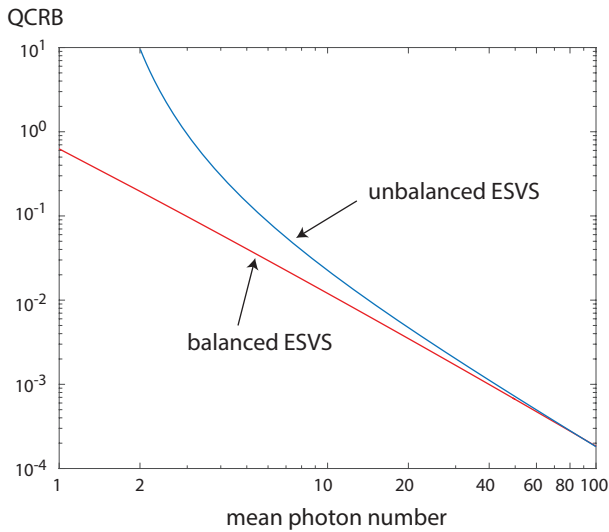}
\caption{Plots of the QCRB using the balanced entangled squeezed vacuum state $|\Psi\rangle_\text{sv}$ (Eq.~(\ref{eq_generalized_state}), red) and the corresponding unbalanced case $|\Psi\rangle_\text{unb}$ (Eq.~(\ref{eq_unbalanced_state}), blue) when $d=5$ with respect to the mean photon number $\bar{n}_\text{sv}$. Note that the mean photon number for the unbalanced squeezed vacuum state with $b = b_\text{bo}$ is always larger than $2$.}
\label{Fig3_QCRB_comparison_b=cVSbnot=c}
\end{figure}

Nevertheless, the QCRB that is optimized with respect to the mode weighting $b$ is not necessarily optimal with respect to the total mean photon number $\bar{n}$, where the latter makes more sense in real experiments.  To illustrate this point, we have chosen the entangled squeezed vacuum state as an example, and plotted the numerical values of $|\delta\boldsymbol{\theta}|_{\text{QCRB}}^2$ using the balanced form $|\Psi\rangle_\text{sv}$ and unbalanced form, respectively. The results are shown in Fig.~\ref{Fig3_QCRB_comparison_b=cVSbnot=c}. Note that for the ESVS unbalanced probe, its QCRB is always minimized at $b_\text{bo}^2$ with respect to $b$.
As we can see, the balanced state can achieve an even lower uncertainty compared to the unbalanced case in the low photon number regime, although their performances are almost identical when the mean photon number is large. Another benefit of the balanced state is its easier experimental implementation because one does not need to distinguish between the reference mode from the probing modes.

%%%%%%%%%%%%%%%%%%%%%%%%%%%%%%%%%%%%%%%%%%%%%%%%%%%%%%%%%%%%%%%%%%%%%%%%%%%%%%%%%%%%%%%%%%%%
\section{Potential Experimental Implementation}

In this section, we present an experimental setup that can produce a two-mode NOON-like state using quantum states that are readily generated as the light sources.  The state can be written as
\begin{equation}
    |\Phi\rangle = \frac{1}{\sqrt{2}} (|\phi\rangle|0\rangle + |0\rangle|\phi\rangle),
    \label{Eq_Experimental_state}
\end{equation}
where $|\phi\rangle = \sum_{n=0}^{4} c_n |n\rangle$ is a superposition of Fock states with up to $4$ photons.

\begin{figure}[!t]
  \centering
  \includegraphics[height=3.3cm]{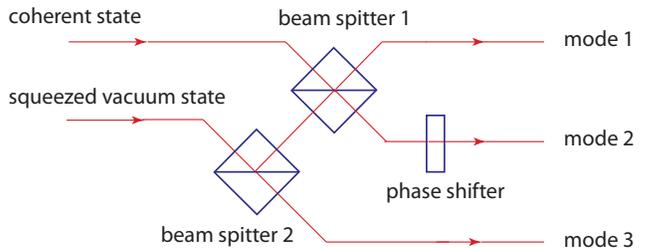}
  \caption{Experimental setup of a two-mode entangled state $|\Phi\rangle$.}
  \label{Fig4_setup}
\end{figure}

The state is produced by combining a coherent state $|\alpha\rangle$ and a squeezed vacuum state $|r\rangle$ with two 50:50 beam splitters, as illustrated in Fig.~\ref{Fig4_setup}. Two conditions are required for the input states: (a) $\alpha$ and $r$ are real values, (b) $\alpha^2=3\tanh{r}/2$.
After the input states propagate through the two beam splitters, a phase shifter with the effective operation $e^{i\pi -\frac{i\pi}{2} a^\dagger a}$ realized by path differences is applied to mode 2, which adds a phase shift linear to the photon number $n$.  At the output, among all the cases of different photon combinations, we post-select the state by detecting one photon in mode $3$ as a trigger, and discard all cases with more than $4$ photons detected in mode $1$ and $2$.
Since current measurement methods are all destructive to photons, the post-selection process would not be applied until the phase sensing is finished.
Then, the state $|\Phi\rangle$ is obtained in mode $1$ and $2$, whose weights for different Fock state terms are
\begin{eqnarray}
    &\displaystyle
    c_0 = 0,
\quad
    c_1 = \frac{ i 2 \sqrt{2}}{g(r)},
\quad
    c_2 = \frac{-2 \sqrt{3} (\tanh{r})^{1/2}}{g(r)},
\nonumber\\
    &\displaystyle
    c_3 = \frac{- i 2 \sqrt{3} \tanh{r}}{g(r)},
\quad
    c_4 = \frac{3 (\tanh{r})^{3/2}}{g(r)},
    &
\end{eqnarray}
where $g(r)=[8+12\tanh{r}+12(\tanh{r})^2+9(\tanh{r})^3]^{1/2}$ is the normalization factor. Note that this state does not contain a vacuum component.

The method described above shares some similarities with the experimental generation of a two-mode NOON state in~\cite{Afek2010_NOON_experiments_cSqV} while there are important differences.  In~\cite{Afek2010_NOON_experiments_cSqV}, pure two-mode NOON states up to 3 photons are produced by combining a coherent state and a squeezed vacuum state using a 50:50 beam splitter.  The NOON state is resulted with the post-selection of the proper subspace that constitutes the desired NOON state.  This method has the advantage that the NOON states are produced by interference effects and no photons are wasted.  It also can generate NOON states of large photon number $N$ with high fidelity though not pure.

On the other hand, the method proposed above has the distinct feature that it is a coherent superposition of Fock states in the form of Eq.~(\ref{Eq_Experimental_state}), whereas one cannot simply mix the different NOON states in~\cite{Afek2010_NOON_experiments_cSqV} as the process is incoherent.  The setup presented above has the advantage of generating a $4$-photon NOON state, which is impossible using the previous work, as well as the potential to be generalized to produce multi-mode states ($d \ge 2$).   Nevertheless, the trigger of single photon in mode 3 results in not using those events with other photon numbers and hence the efficiency of this method is considerably low.

\begin{figure}[!t]
  \centering
  \includegraphics[width=8cm]{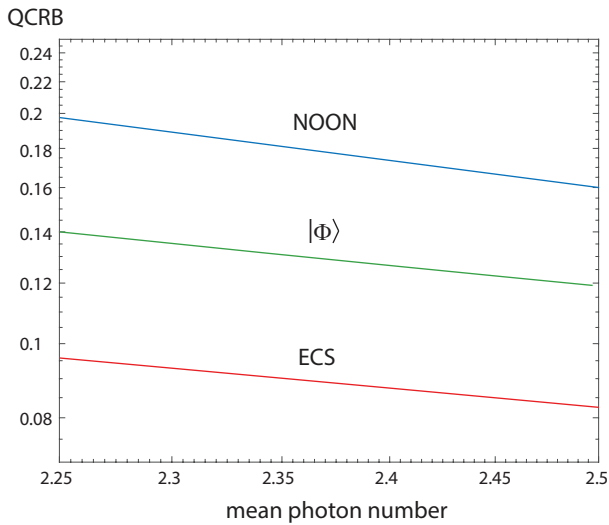}
  \caption{Plots of the QCRB with the NOON state (blue), ECS (red), and $|\Phi\rangle$ (green). Note that the plot for the NOON state corresponds to the effective situations with the non-integer photon number for comparisons.}\label{Fig5_com_exp}
\end{figure}

To show the performance of $|\Phi\rangle$ in quantum phase imaging, we plot its QCRB numerically, along with the cases of the NOON state and the ECS, within the region $2.25\leq \bar{n} \leq 2.5$, as shown in Fig.~\ref{Fig5_com_exp}. This region is achieved when $1\leq r \leq 2$, which is chosen in order to obtain the optimal generation probability.  It is seen that this state can perform better than the NOON state ``effectively'' with the same mean photon number, but worse than the entangled coherent state.

%%%%%%%%%%%%%%%%%%%%%%%%%%%%%%%%%%%%%%%%%%%%%%%%%%%%%%%%%%%%%%%%%%%%%%%%%%%%%%%%%%%%%%%%%%%%%
\section{Summary}

In this paper, the simultaneous multi-parameter estimation problem is investigated using a multi-mode NOON-like entangled state, which is formed by replacing the non-vacuum mode (i.e., the Fock state) in NOON state with an arbitrary state.  The analytical form of the QCRB using this state is presented.  It shows that the QCRB is a monotonic increasing function of the parameter $f \equiv \langle \hat{H} \rangle / \langle \hat{H}^2 \rangle$, where $\hat{H}$ is the interaction Hamiltonian of the phase estimation problem, in addition to its dependence on the number of modes $d$ and the mean photon number $\bar{n}$.  Interestingly, we proved that the NOON state has the worst performance among the whole class of states given fixed $d$ and $\bar{n}$.

As specific examples, we studied the estimation performances when the probe state is a multi-mode NOON state, entangled coherent state, entangled squeezed coherent state, and entangled squeezed vacuum state.  Through the comparisons among these different probe states, we found that with the same mean number of photons, the entangled squeezed vacuum state has the lowest estimation uncertainty and hence leads to the best multiple phase estimation precision.  For the squeezed coherent state scenario, the uncertainty decreases as the squeeze factor increases.  In addition, we show that for the squeezed vacuum case, the presented state has a better performance than its unbalanced counterpart adopted in the previous works.  Finally, we described an experimental setup to produce an example of a generalized two-mode entangled state that can beat the NOON state with respect to the QCRB.

%%%%%%%%%%%%%%%%%%%%%%%%%%%%%%%%%%%%%%%%%%%%%%%%%%%%%%%%%%%%%%%%%%%%%%%%%%%%%%%%%%%%%%%%%%%%%%%%%%%%%%%
\appendix

\section{Proof of Eq.~(\ref{eq_QCRB_comparison_noon_coh_sc_sv})}

Equation~(\ref{eq_QCRB_function_f}) gives a general formula of the QCRB for any state with the form as in Eq.~(\ref{eq_generalized_state}). As we state in the main text, $|\delta\boldsymbol{\theta}|_{\text{QCRB}}^2$ is a monotonic increasing function with respect to the parameter $f \equiv \langle\hat{H}\rangle/\langle \hat{H}^2 \rangle$ when the phase dimension $d$ and the mean total photon number $\bar{n}$ are fixed. Due to the variance $(\langle \hat{H}^2 \rangle - \langle\hat{H}\rangle^2) \geq 0$, we can get $0 \leq f \leq 1/\bar{n}$, where the upper bound is derived as below:
\begin{equation}
    f
    =
    \frac{\langle\hat{H}\rangle}{\langle \hat{H}^2 \rangle}
    \leq
    \frac{\langle\hat{H}\rangle}{\langle \hat{H}\rangle^2} = \frac{1}{\tilde{n}} = \frac{1}{\bar{n}(1+d|\langle \psi|0\rangle|^2)}
    \leq
    \frac{1}{\bar{n}}.
\label{f}
\end{equation}
The upper bound is saturated when $\langle \hat{H}^2 \rangle = \langle \hat{H}\rangle^2$ and $|\langle \psi|0\rangle|^2=0$, i.e., NOON state, which leads to Eq.~(\ref{eq_QCRB_ineq}).

In order to obtain the QCRB inequality between the four scenarios as shown in Eq.~(\ref{eq_QCRB_comparison_noon_coh_sc_sv}), we only need to compare their corresponding $f$ factors:
\begin{eqnarray}
    &\displaystyle f_N = \frac{1}{\tilde{n}_N},
    \ \ \ \  f_\text{c} = \frac{1}{\tilde{n}_\text{c}+1},
\nonumber\\
    &\displaystyle f_\text{sc} = \left(\tilde{n}_\text{sc}+\frac{\alpha'^2 e^{2r'}+ 2\sinh^2{r'}\cosh^2{r'}}{\alpha'^2+\sinh^2{r'}}\right)^{-1},
\nonumber\\
    &\displaystyle f_\text{sv} = \frac{1}{\tilde{n}_\text{sv}+2\cosh^2{r}},
\label{4f}
\end{eqnarray}
where the mean photon numbers $\tilde{n}_N$, $\tilde{n}_\text{c}$, $\tilde{n}_\text{sc}$ and $\tilde{n}_\text{sv}$ for the non-vacuum modes  are defined in the same ways as in the main text.
When the mean total photon numbers are fixed (i.e., $\bar{n}_N=\bar{n}_\text{c}=\bar{n}_\text{sc}=\bar{n}_\text{sv}$), we can easily derive
\begin{equation}
    \tilde{n}_N < \tilde{n}_\text{c} < \tilde{n}_\text{sc} < \tilde{n}_\text{sv}
    \label{A_eq_nbar_4}
\end{equation}
from Eq.~(\ref{eq_mean_photon_number_4}) and using some fundamental mathematical calculations.
Under the conditions of Eq.~(\ref{A_eq_nbar_4}) and using the properties of hyperbolic functions, we can obtain
\begin{equation}
    1  < \frac{\alpha'^2 e^{2r'}+ 2\sinh^2{r'}\cosh^2{r'}}{\alpha'^2+\sinh^2{r'}} < 2\cosh^2{r}.
    \label{A_eq_other_4}
\end{equation}
From Eq.~(\ref{A_eq_nbar_4}) and Eq.~(\ref{A_eq_other_4}), we can derive the inequalities $f_N > f_{\text{c}} > f_{\text{sc}} > f_{\text{sv}}$, which in turn give Eq.~(\ref{eq_QCRB_comparison_noon_coh_sc_sv}).

\end{document}